# Sub-micron single-particle perovskite plasmonic nanolasers at room temperature

Sangyeon Cho[1,2], Yi Yang[3], Marin Soljačić[3], Seok Hyun Yun[1,2]

[1]Wellman Center for Photomedicine, Massachusetts General Hospital and Harvard Medical School, 65 Landsdowne St., Cambridge, Massachusetts, 02139, USA

[2]Harvard-MIT Health Sciences and Technology, Massachusetts Institute of Technology, 77 Massachusetts Avenue, Cambridge, Massachusetts, 02139, USA

[3]Research Laboratory of Electronics, Massachusetts Institute of Technology, 77 Massachusetts Avenue, Cambridge, Massachusetts, 02139, USA

**Abstract**

Plasmonic nanolasers have received a substantial interest for their promising applications in integrated photonics, optical sensing, and biomedical imaging. To date, a room-temperature plasmonic nanolaser, sub-micron in all dimensions, remains elusive in the visible regime due to high metallic losses. Here, we demonstrate single-particle lasing around 2.3 eV with full-submicron, cesium lead bromide perovskite ($CsPbBr_3$) crystals atop polymer-coated gold substrates at room temperature. With a large number (~100) of devices in total, we systematically study the lasing action of plasmonic test and photonic control groups. The achieved smallest plasmonic laser was 0.56 μm × 0.58 μm × 0.32 μm in size, ten-fold smaller than that of our smallest photonic laser. Key elements to efficient plasmonic lasing are identified as enhanced optical gain by the Purcell effect, long carrier diffusivity, a large spontaneous emission factor, and a high group index. Our results shed light on three-dimensional miniaturization of plasmonic lasers.

**Main Text**

Nanoscale lasers have received a growing interest for scientific and practical applications (*1–3*). Incorporating metals, which support highly-confined surface-plasmon polaritons (SPPs), has become a promising route to reduce device sizes beyond what is possible with only semiconductors and dielectrics. One class of metal-based lasers, known as spaser, relies on localized surface plasmon resonances. Since its inception (*4*), a limited number of demonstrations (*5–8*) have been reported, although the samples under study are in the form of colloidal ensembles instead of single particles (*9*). Another class of metal-based lasers, known as plasmonic nanolasers, relies on propagating SPPs. These plasmonic nanolasers have been demonstrated using various gain materials and metals (*10–13*). In the near-infrared regime, full submicron nanolasers in all dimensions have been realized with metallic-coated disks (*14*). Nevertheless, this goal becomes more challenging in the visible regime because of the relatively low gain in semiconductors (*15*) and the more absorptive response of metals. Although ultra-thin geometries, such as nanowires and microplates, are possible, at least one dimension along the cavity oscillation direction has to be micron-sized to compensate for propagation losses (*16, 17*). To date, in the visible regime, a room-temperature plasmonic nanolaser with full-submicron sizes in all three dimensions has remained experimentally elusive (*18*).

Lead halide perovskite (LHP) with the form of $APbX_3$ ($A=Cs^+$, $CH_3NH_3^+$, $X=Cl^-$, $Br^-$, $I^-$) have emerged as solution-processable semiconductor materials for high-performance optoelectronic applications (*19–22*). Among different LHPs, cesium lead bromide perovskites ($CsPbBr_3$) is an excellent laser material owing to its long carrier diffusion length (~9 μm) (*23*) and high optical gain (*24*): e.g. 4000 $cm^{-1}$ near 2.3 eV. Such extraordinary properties result from a combinatory effect of large active carriers by Coulomb correlated electron-hole plasma (EHP) (*25, 26*) and large polaron formation by deformation of $PbBr_3^-$ sublattice (*27*). Recent studies suggest that $CsPbBr_3$ lasing occurs in a state of quasi-three-level EHP above Mott density ($\rho_{mott}= 10^{17-18}$ /$cm^3$) (*26, 28*), which is beneficial to build up a large population inversion by band gap renormalization (BGR) (*29*).

Here, we demonstrate full-submicron perovskite plasmonic nanolasers in the visible regime with nanosecond optical pumping at room temperature. Our laser devices, solely fabricated with solution-based chemistry, harness $CsPbBr_3$ submicron nanocubes atop a polymer-coated gold substrate. We achieve stable single- and multi-mode lasing around 2.3 eV (=540 nm) with a narrow laser linewidth of 1 meV (=0.2 nm) under a threshold fluence around 0.3 mJ/$cm^2$. By measuring a number of laser parameters, including the Purcell factor, the spontaneous emission factor (*β*), and the quantum yield (*η*), with spectroscopic tools, we systematically study the lasing action of both plasmonic lasers (test) and their photonic counterparts (control). Such a comparison enables us to identify and explain a condition where efficient plasmonic lasing becomes possible.

We synthesized $CsPbBr_3$ micron- and submicron-sized crystals using a sonochemistry method (*30*) (Fig. S1). Prepared $CsPbBr_3$ crystals were subsequently transferred onto different substrates: gold (Au) coated with polyepinephrine (pNE), pNE-coated silicon (Si), and bare silica ($SiO_2$) (Fig. 1A). $CsPbBr_3$ crystals on the gold substrate form the plasmonic laser group (test), while those on dielectric $SiO_2$ or Si substrates are the two photonic counterparts (control). The quantities of devices in the test and control groups are 30 (Au), 25 (Si), and 45 (SiO$_2$), respectively. In the plasmonic test group, the thickness of the pNE spacer is chosen as 5 nm, which enabled the smallest lasers in our experiment (see the comparison of laser performances at different spacer thicknesses 1 nm, 5 nm, and 10 nm in Fig. S2).

We pumped every single device with nanosecond optical pulses at 480 nm (duration 4 ns, repetition rate 20 Hz). For each device, we recorded output emission spectra at various pump levels (Fig. S3) below and above the lasing threshold. Their locations on the substrate were subsequently marked (Fig. S4) by pump laser irradiation at high fluence for scanning electron microscopy (SEM) imaging to determine the device sizes. The pump fluences $E_{th}$ at the lasing threshold of the three experimental groups are shown in Fig. 1B as a function of the effective device side length $L = \sqrt{S}$, where $S$ is the device area in the $x$-$y$ plane. A large number of plasmonic lasers achieved lasing with submicron sizes. Ten smallest plasmonic lasers are denoted by $i$ to $x$ in Fig. 1B.

The smallest plasmonic laser (Fig. 2A, inset) has $L$=0.57 μm ($i$; 0.56 μm × 0.58 μm × 0.32 μm for $x$, $y$, and $z$ directions), which is about 10-fold smaller in device volume $V$ than that of the smallest photonic laser. Besides, the second smallest device ($ii$; 0.75 μm × 0.49 μm × 0.3 μm; $L$=0.61 μm) showed single-mode lasing (Fig. 2B, inset). The gold substrate facilitated substantial reduction of the device sizes as no crystals smaller than $L \approx 1.2$ μm on the dielectric substrates showed lasing. Regardless of $L$, the lasing thresholds were similar across all the three groups (Fig. 1B). Below the lasing thresholds, the plasmonic devices, regardless of sizes, generally exhibited multiple peaks in their PL spectra, indicating increased density of states (Figs. 2A, 2B, and S5). Above the thresholds, narrow emission peaks appear with substantial increase of intensity. The device $i$ exhibited two lasing modes, M1 and M2 (Fig. 2A). M1 reached lasing first, followed by M2. The device $ii$ demonstrated a single lasing mode. (Fig. 2B). We observed an immediate reduction of the full-width-half-maximum (FWHM) of the PL emission peaks near the thresholds in both devices using the multi-peak Lorentzian fitting method (Figs. 2C, 2D, and S6). The narrowest FWHM of the lasing peaks near the threshold is about 2.4 meV (=0.7 nm) and 2.2 meV (=0.5 nm) for both M1 and M2 in device $i$, and 1 meV (=0.2 nm) in device $ii$. Both the linewidth narrowing and a polarization change (Fig. S6E) are clear indications of lasing. In contrast, all photonic lasers showed broad single-peak profiles in the PL spectra below the lasing threshold (Figs. 2E, S6G, and S7).

From hyperspectral images (Fig. S8), we reconstructed the stimulated emission profiles of both M1 and M2, which were revealed to be Fabry-Perot and whispering-gallery modes (WGM), respectively (Fig. 2F). We numerically calculated these two eigenmodes (COMSOL Multiphysics). The simulated quality factors are 61 and 41 for M1 and M2, respectively. Both modes are hybrid resonances, because their electric fields are concentrated inside the gap region and also oscillate within the CsPbBr$_3$ crystal (Fig. 2G). Conceptually, these resonances can be understood as the surface plasmons of a planar multilayer

metal−dielectric closed system restricted to specific quantized wavevectors determined by the device sizes (*31*). We calculate the plasmon dispersions of such closed systems where the CsPbBr$_3$ layer is infinitely extended in the $x$ and $y$ directions (Fig. S9). From the dispersion, group indices $n_g$ of the two modes are 2.95 (M1) and 4.1 (M2), and their propagation losses α are 1,800 cm$^{-1}$ (M1) and 5,800 cm$^{-1}$ (M2). In the photonic control group with silicon substrates, two photonic modes are identified with $Q_m \approx 90$ and 38, respectively, and $n_g \approx 2.5$ (Fig. S9). Therefore, in our devices, the plasmonic system can achieve similar quality factors to those of the photonic system. This is because the higher $n_g$ of the plasmonic modes results in stronger SPP boundary reflections circumventing the CsPbBr$_3$ crystal, which reduces the radiation losses, and compensates the increased absorption losses.

We studied the laser dynamics with a three-level rate equation model (*16*). In this model, the threshold pump absorption rate $P_{th}$ per volume $V$ at frequency $\omega_0$, (Supplementary Note 1) is given by $P_{th}V \approx \omega_0/Q_m\beta_m\eta$. Here, $\beta_m$ is the spontaneous emission factor of the lasing mode $m$, and $\eta$ is quantum yield of the gain medium. $P_{th}$ was calculated from the experimentally measured threshold pump fluence $E_{th}$, ranges from 0.5 to 5 × 10$^{18}$ cm$^{-3}$ns$^{-1}$ for all lasing devices (Fig. S10). To reduce the lasing threshold, three parameters in the denominator are important: quality factor $Q_m$ of the lasing mode, spontaneous emission factor $\beta_m$ of the lasing mode, and the quantum yield $\eta$ of the gain medium. Our simulation results above have predicted similar quality factors in our plasmonic and photonic devices. Below, we show that plasmonic devices, compared to the photonic test groups, offer higher $\beta_m$ and $\eta$ for submicron $V$.

We used both laser experiment data and transient spectroscopic tools to analyze a few important laser parameters of our devices, including $\beta_m$, $\eta$, and Purcell factor. First, we extracted $\beta_m$ of the lasing mode of the plasmonic devices at a fixed, low pump fluence at 0.2 mJ/cm$^2$ (Fig. 3A). We employ an analytical ray tracing model (*12*, *32*) to calculate the number of cavity modes ($N$) (Fig. S12), which varies between 3 and 6 within the gain bandwidth. For the plasmonic devices, we obtain $\beta_m$ by decomposing the PL spectra, which contain multiple emission peaks, into $N$ Lorentzians and background fluorescence for uncoupled emission (Fig. S6). We observe that the plasmonic $\beta_m$ increases from ~0.05 to ~0.35 as the device size decreases from ~1.3 μm to ~0.6 μm. $\beta_m$ in photonic devices cannot be inferred from their PL spectra because of the lack of multiple resonant peaks (Fig. S7).

Next, we probed the spontaneous emission enhancement with two approaches, the laser experiments (Fig. 3B) and the time-resolved spectroscopy (Fig. 3C). In an integrating sphere measurement, we benchmarked the intrinsic quantum yield ($\eta_{int}$) ≈ 1.5% for our synthesized CsPbBr$_3$ microcrystals (> 5 μm), synthesized in a N,N-dimehtlformaldehyde (DMF) solution and transferred to Si substrates. Compared to this benchmark, spontaneous emission at the same effective pump fluence (~0.2 mJ/cm$^2$) was substantially enhanced in the plasmonic devices compared to that in the photonic ones, as shown in our laser experiments (Fig. 3B). In particular, the plasmonic devices, especially those submicron sized, were 10-40x brighter. Consequently, the $\eta$ of the gain media in the plasmonic devices were increased to 0.15-0.6, an order of magnitude higher than those in the photonic devices. Spontaneous emission enhancement (the Purcell effect) via SPPs in the vicinity of plasmonic cavities has been well studied (*33, 34*). For the measured $\beta_m$ and $\eta$ of various devices, we devised an analytical model to explain their general trends with a decent agreement achieved (Fig. S13).

In our second approach, we verified such emission enhancement using time-resolved spectroscopy. We show representative fluorescence decays of $CsPbBr_3$ crystals of three different sizes on pNE coated Au substrates (Fig. 3C). More comprehensive measurements on more samples on both Au and Si substrates are shown in Fig. S14 and Table S1. We observed that the peak intensities and the total number of emitted photons, proportional to the total spontaneous emission enhancement (Supplmentary Note 1D), are simultaneously increased for small devices. From these data, we extracted the associated total Purcell factor and $\eta$ (Fig. S14 and Table S1), which are at least one order of magnitude higher than those in the photonic control group. These results are in agreement with those obtained from the previous alternative methods (Figs. 3A and 3B).

The light-in-light-out (L-L) curve is another crucial indicator of laser performance. Specifically, the nonlinear kink in the L-L curves is determined by the parameter $x \equiv \beta_m \eta$ (*35*) (Fig. 3D and Supplementary Note 1B). Figure 3D shows the L-L curves along with the extracted *x* parameters of our three devices. The extracted *x* parameters agree reasonably well with the product of the spontaneous emission factor of the lasing mode $\beta_m$ and quantum yields $\eta$ that are obtained independently in Figs. 3A and 3B.

Our systematic study of the lasing actions in plasmonic and photonic devices enables us to identify a regime where efficient plasmonic lasing becomes possible. As shown in Fig. 4, we plot the measured threshold carrier density $\rho_{th}$, inferred from $P_{th}$ (Supplementary Note 1C), versus the device sizes in all the three experimental groups. We also plot the associated theoretical calculations (solid and dashed lines) via $\rho_{th} V \approx \omega_0 \tau_s / Q_m \beta_m F_{tot}$ (Supplementary Note 2) for our lasing devices around 2.3 eV. The comparison between the theoretical curves shows that the plasmonic (photonic) devices exhibit lower lasing thresholds for small (large) devices. The crossover of the two curves is around 2 μm. These predictions agree well with our measurements (squares). The increased group indices and Purcell factors in the plasmonic devices compensate for the absorptive losses and thereby reduce the lasing thresholds (Supplementary Note 2). Therefore, more efficient plasmonic lasing becomes possible, as indicated by the shaded yellow region in Fig. 4.

To achieve efficient plasmonic lasing, CsPbBr$_3$ has two unique advantages, high carrier density and long carrier diffusion length (> 9 μm). From a material perspective, the maximal attainable $\rho$ (proportional to optical gain) is limited by various loss mechanisms, such as Auger recombination and heat-induced material damage, even under strong pumping. In CsPbBr$_3$ near the lasing threshold, carrier density $\rho$ exceeds the Mott transition density $\rho_{Mott}$ (dashed green line in Fig. 4) and reaches $10^{18}$-$10^{19}$ cm$^{-3}$, forming electron-hole plasma (EHP)(*26*) (Supplementary Note 3). At $\rho = 10^{19}$ cm$^{-3}$, the gain of bulk CsPbBr$_3$ is measured as about 4000 cm$^{-1}$ (*24*). Such high gain around 2.3 eV (*28*) is unique compared with other gain materials, including organic materials (e.g. poly(phenylene vinylene) polymers, ~90 cm$^{-1}$ (*36*)) or inorganic III/V semiconductor (e.g. InGaN quantum wells, 1500 cm$^{-1}$ (*37*)). Another advantage is that the diffusion length of charge carriers in CsPbBr$_3$ exceeds 9 μm (*23*, *38*). The charge carriers created over the entire device volume can migrate and undergo Purcell-enhanced radiative recombination upon increased overlap with plasmonic lasing modes. In this sense, gain media with low carrier diffusivity would be far less desirable for efficient plasmonic lasing.

We have demonstrated CsPbBr$_3$-on-gold nanolasers operated in the efficient plasmonic lasing regime. Compared to previous works (Table S2) in the visible wavelengths, the realized devices further reduce the length scales of room-temperature plasmonic nanolasers to the full-submicron regime. Furthermore, our devices, including both the semiconductor gain material and the polymer coating, are fabricated with solution-based chemistry only. Therefore, our techniques hold unique promise for mass production of high-performance nanolasers at low cost. The intrinsic $\eta$ of our fabricated devices could be improved by defect reduction, as near 100% $\eta$ in CsPbBr$_3$ quantum dots has been demonstrated (*39*). Besides, lasing thresholds could be further reduced with Fano resonances in optimized structures (*40*, *41*). These combinatory efforts may further miniaturize plasmonic nanolasers towards the deep sub-diffraction limit.

**Acknowledgements**

The authors thank Kwon-Hyun Kim for spectroscopic ellipsometry measurement. This research was supported by National Institutes of Health (grant no. DP1EB024242) and the Massachusetts General Hospital Research Scholar Award. S.C. acknowledges the Samsung Scholarship. Part of this work used the facilities in the Center for Materials Science and Engineering at MIT and the Center for Nanoscale Systems, a part of Harvard University and a member of the National Nanotechnology Coordinated Infrastructure Network (NNCI), which is supported by the NSF (grant no. 1541959). This material is based upon work supported in part by the U.S. Army Research Office through the Institute for Soldier Nanotechnologies at MIT, under Collaborative Agreement Number W911NF-18-2-0048.

**Author contributions**

S.H.Y. and S.C. conceived and designed the project. S.C. performed the experiments and analyzed the data. Y.Y. provided theoretical assiatance. All authors contributed to manuscript writing. M.S. and S.H.Y. supervised the project.

**Competing interests**

The authors declare no competing interests.

**Corresponding author**

Correspondence to S.H.Y (syun@hms.harvard.edu)

**Images with captions**

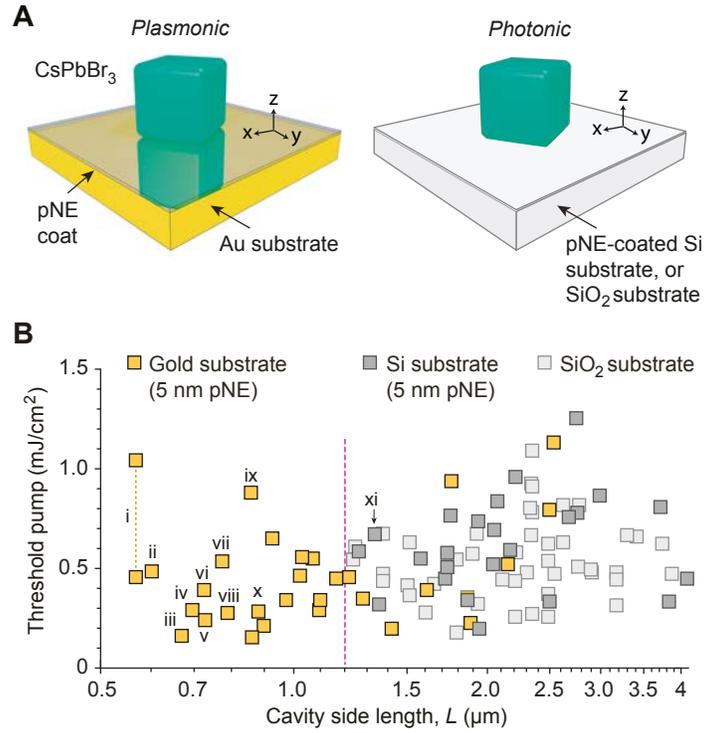

**Fig. 1. Plasmonic versus photonic lasers. A**, Schematic of plasmonic (polynorepinephrine (pNE)-coated Au substrate) and photonic (pNE-coated Si or bare SiO$_2$ substates) devices. **B**, Threshold pump fluence $E_{th}$ measured from individual lasers. Substantial reduction of device sizes is observed with the plasmonic devices. The number of devices are 30, 25, and 45 on Au, Si, and SiO$_2$ substrates, respectively. Ten submicron plamonic lasers are labeled as *i* to *x*, and one photonic laser is marked *xi*.

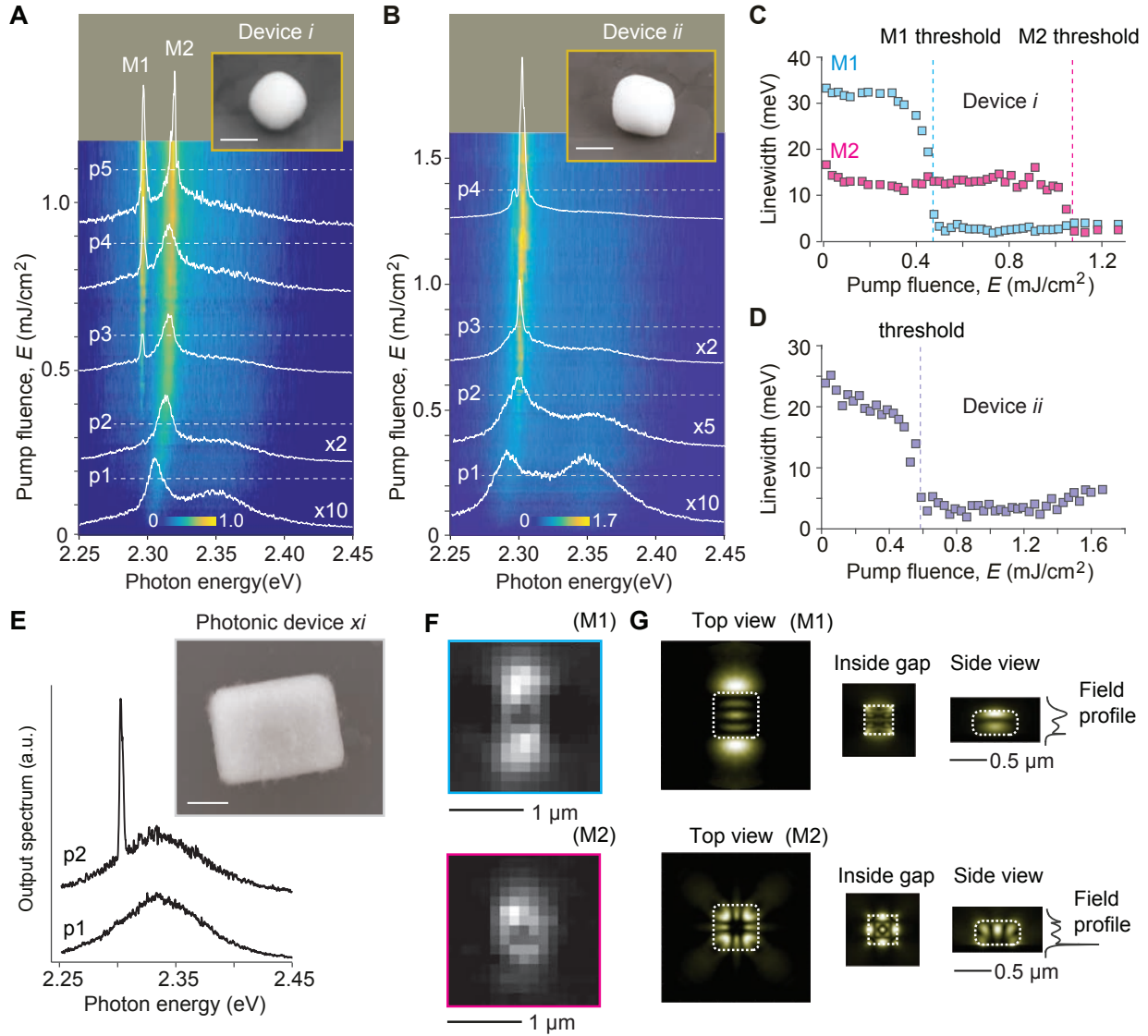

**Fig. 2. Emission spectra and field profiles. A-B**, Single-shot photonluminescence (PL) spectra of the two smallest plasmonic lasers (Device *i*, *L*=0.57 μm) (A) and (Device *ii*, *L*= 0.61 μm) (B) and their line cuts at various pump fluences. Two lasing modes in device *i* are denoted by M1 and M2. Scale bars, 500 nm. **C-D**, Measured spectral linewidth of device *i* (C) and device *ii* (D) at different pump fluences. There are clear linewidth reductions at the lasing thresholds. **E**, Single-shot PL spectra from a photonic device *xi* (*L*=1.3 μm). Scale bar, 500 nm. **F**, Measured wide-field images of the stimulated emission (see Fig. S8) of device *i*. **G**, Intensity profiles of the calculated cavity modes, revealing the Fabry-Perot and whispering-gallery nature of the two hybrid plasmonic modes in device *i*. White dotted lines indicate device boundaries.

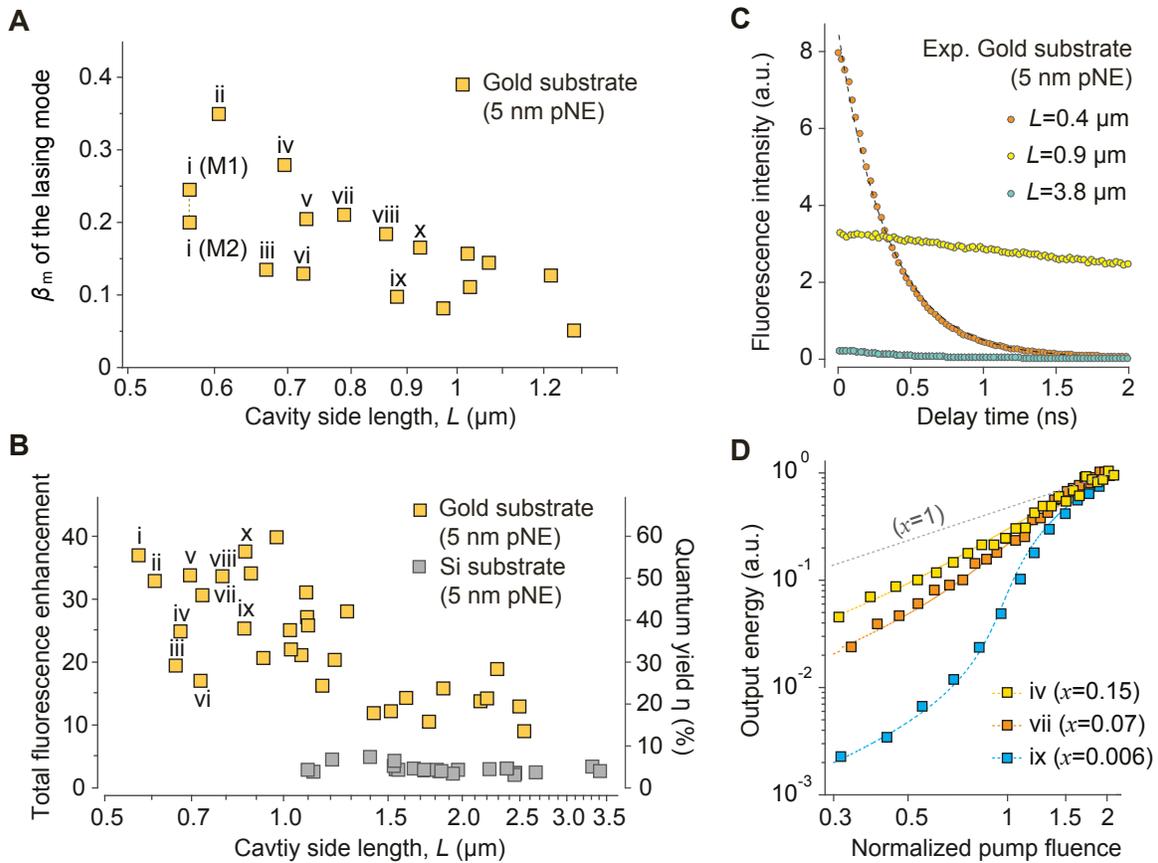

**Fig. 3. Laser Characteristics. A**, Extracted $\beta$-factor of the lasing modes for the plasmonic devices (yellow squares). **B**, Measured fluorescence enhancement (10-40 times) and the associated quantum yields $\eta$ of the plasmonic (yellow; polynorepinephrine (pNE)-coated Au) and photonic (grey; pNE-coated Si) devices. **C**, Time-resolved photoluminescence decay (circles) of three $CsPbBr_3$ plasmonic devices and their double-exponential fits (dashed lines). The smallest device exhibits enhanced fluorescence intensity and accelerated decay rates. **D**, Measured (squares) Light-in-Light-out (LL) curves of three plasmonic devices *iv* ($\beta_m$=0.28, $\eta$=0.37), *vii* ($\beta_m$=0.21, $\eta$=0.46), and *ix* ($\beta_m$=0.1, $\eta$=0.38) and the associated fits (dashed lines; See Supplementary Note 1B).

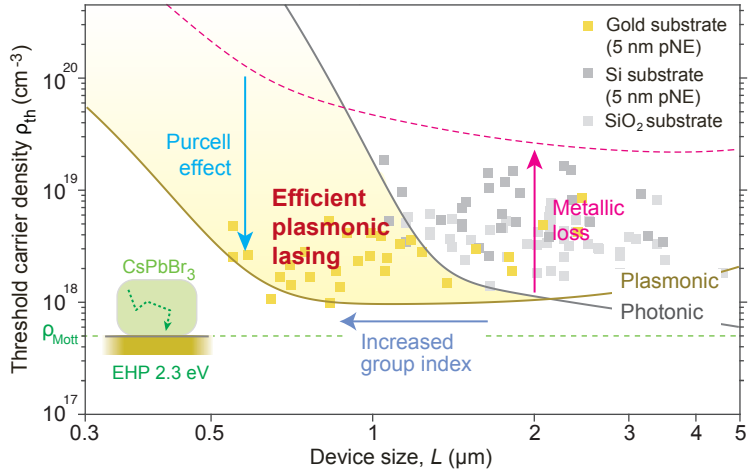

**Fig. 4. Efficient plasmonic lasing from hybrid modes of submicron CsPbBr₃ particles.** As laser device size decreases, an efficient plasmonic lasing regime (yellow region) emerges. This is indicated by a crossover around 2 μm in the calculated (see Supplementary Note 2) threshold carrier density for our plasmonic (brown solid line) lasers and photonic (grey solid line) operated at 2.3 eV, consistent with our experimental data (squares; calculated from Fig. 1B; see Fig. S10). Metallic loss (upward arrow) increases the threshold (pink dashed curve), which is compensated by the Purcell effect (downward arrow). The increased group index reduces simultaneously the device size (leftward arrow) and the threshold by supressing the radiative loss. Inset: device schematic illustrating the migration of charge carriers in the strongly pumped electron hole plasma (EHP) state above the Mott transition density ($\rho_{Mott}$) of CsPbBr₃ (dashed green line).

# Supplementary Information for

## Sub-micron single-particle perovskite plasmonic nanolasers at room temperature

Sangyeon Cho[1,2], Yi Yang[3], Marin Soljačić[3], Seok Hyun Yun[1,2]

[1]Wellman Center for Photomedicine, Massachusetts General Hospital and Harvard Medical School, 65 Landsdowne St., Cambridge, Massachusetts, 02139, USA

[2]Harvard-MIT Health Sciences and Technology, Massachusetts Institute of Technology, 77 Massachusetts Avenue, Cambridge, Massachusetts, 02139, USA

[3]Research Laboratory of Electronics, Massachusetts Institute of Technology, 77 Massachusetts Avenue, Cambridge, Massachusetts, 02139, USA

## Methods

Fabrication of laser devices. For producing $CsPbBr_3$, $CsBr$ and $PbBr_2$ were dispersed at an equal saturating concentration (typically 75 mM each) in N,N-dimethylformamide (DMF) in a vial. The vial was then sonicated at 20-80 kHz in a bath-type ultrasonicator or a tip ultrasonicator in room temperature. After 2-3 min of ultrasonication, single-phase $CsPbBr_3$ micron- and submicron-sized crystals are spontaneously formed. For polynorepinephrine (pNE) coating, 2 mg of (±)-norepinephrine (+)-bitartrate salt were mixed with phosphate-buffered saline (PBS) buffer, and pH was adjusted to 8.5. A polycrystalline Au film-coated substrate (Playtypus) was placed in this solution at room temperature. An incubation time of 2.5 hr produces a 5 nm-thick pNE layer on the gold. The coated film was cleaned and dried under a stream of $N_2$ flow. Finally, the prepared $CsPbBr_3$ crystals were drop-casted on pNE-coated Au substrates. For photonic devices, pNE-coated Si substrates or uncoated $SiO_2$ substrates were used.

Laser characterization. Lasing is observed in ambient air. The samples are placed in a home-built epi-fluorescence microscopy setup (Fig. S3). The pump light source was an optical parametric oscillator (OPO, Optotek HE 355 LD) tuned to 480 nm with a repetition rate of 20 kHz and a pulse duration of 4 ns. The pump light in a circular polarization state was focused to a single device via a 0.5-NA, 50x air objective lens (Nikon) with a full-width-at-half-maxima beam width of ∼ 25 μm. The output emission from the device is collected by the objective lens, passed through a dichroic mirror and a dichroic filter, and split to an EMCCD camera (Luca, Andor) for wide-field imaging and to a grating-based EMCCD spectrometer (Shamrock, Andor) with a spectral resolution of ∼ 0.1 nm. For absolute quantum yield measurement, we used a continuous-wave laser at 491 nm (Cobolt Calypso) for excitation. For time-resolved PL measurements, we used a picosecond laser (VisIR-765, PicoQuant), which was frequency-doubled to 382 nm for excitation, a single-photon avalanche photodiode (Micro Photonics Devices) with a response time of 50 ps, and a time-correlated single-photon counting board (TimeHarp 260, PicoQuant) with a resolution of 25 ps. All optical measurements were conducted at room temperature.

**FDTD simulations**. The eigenmodes and field profiles of the submicron lasers were numerically simulated with a commercial finite-element solver COMSOL Multiphysics with geometries as close to the shapes found by scanning electron microscopy as possible. The Purcell factors were obtained by calculating the spontaneous emission rates of vertical-oriented dipoles inside the gap region.

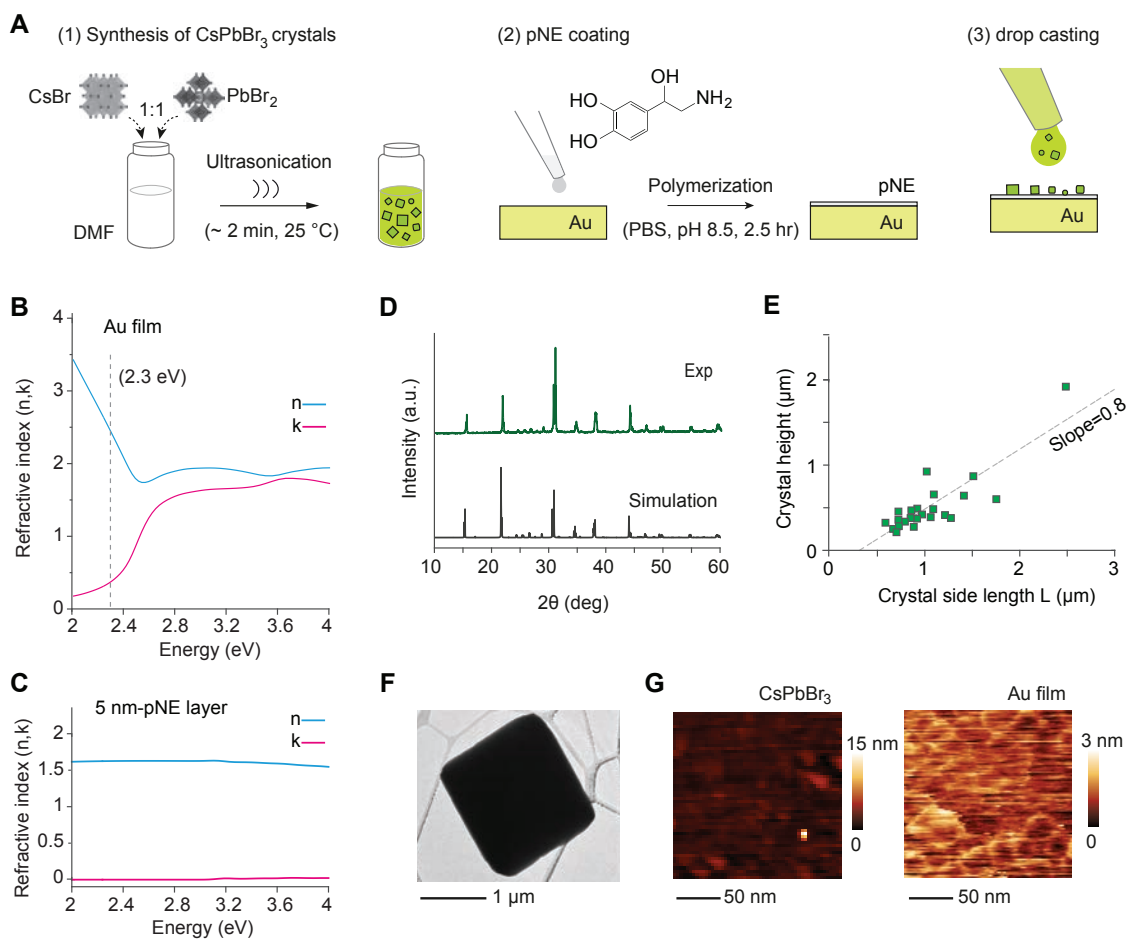

**Fig. S1. Fabrication and properties of CsPbBr₃ crystal, pNE layer, and Au film. A**, Fabrication process. (**B-C**) Complex refractive indices (n, k) of bare Au film (**B**) and 5-nm thick pNE layer-coated on Au film (**C**) measured using spectroscopic ellipsometry. The Tauc-Lorentz dispersion model was used to fit the measured ellipsometry spectra. **D**, Measured and simulated powder X-ray diffraction patterns assuming orthorhombic CsPbBr₃ (space group *Pbnm*, $a$ = 8.20 Å, $b$ = 8.24 Å, $c$ = 11.74 Å). **E**, Aspect ratios of CsPbBr₃ crystals measured using SEM. The linear slope of 0.8 is consistent with the orthorhombic crystal structure. **F**, TEM image of a CsPbBr₃ crystal, revealing smooth surface. **G**, AFM images of a CsPbBr₃ microcube and gold film, showing a root-mean-square roughness of ~ 2.1 nm for the gain crystal and ~ 0.7 nm for the gold surface.

Methods: The complex refractive index was measured from variable angle spectroscopic ellipsometry data measured over 55° to 75° in steps of 5° (J. A. Woollam V-VASE32). The spectral range was from 300 nm to 620 nm. The optical constants fitting was initiated with the Cauchy model at the transparent region and completed by the Gaussian oscillator model. For PXRD, data over 2θ angles from 10° to 60° were collected using a PANalytical X'Pert PRO high-resolution X-ray diffraction system with a CuKα irradiation source. For SEM and EDX, samples were transferred onto a chipped Si wafer by drop casting and imaged using a Zeiss Merlin high-resolution SEM equipped with an EDX detector operated at 15 kV. For TEM, samples were drop casted onto a grid (Ted Pella), and images were acquired using a FEI Tecnai Multipurpose TEM at 120 kV. Illumination beam was expanded to avoid sample damage. AFM images were acquired using a Nanoscope IV Scanning Probe Microscope (Veeco Metrology Group) in a tapping mode. Spectroscopic ellipsometry was demonstrated at the Harvard Center for Nanoscale Systems. All other measurements were performed at the MIT Center for Material Science and Engineering.

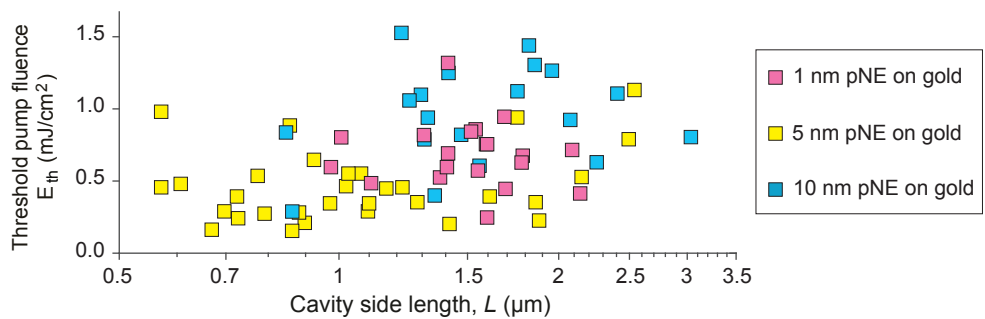

**Fig. S2. Optimization of pNE thickness.** Measured threshold pump fluence of plasmonic lasers fabricated with a different thickness of pNE layer. The 5-nm thickness produces the best result in terms of device size.

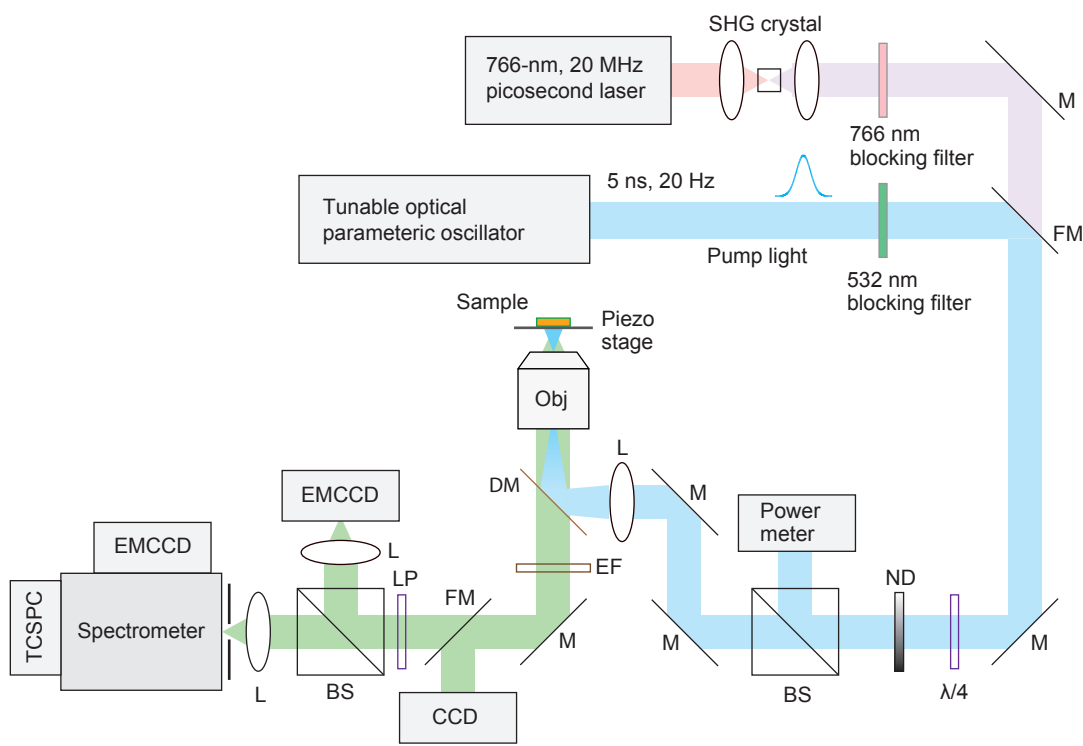

**Fig. S3.** Schematic of the optical characterization setup. L: lens, SHG: second-harmonic generation, M: mirror, FM: flip mirror, LP: linear polarizer, DM: dichroic mirror, ND: neutral density filter, BS: beam splitter, EF: emission filter, CCD: charge-coupled device camera, EMCCD: electron-multiplication CCD camera, and TCSPC: time-correlation single-photon counter.

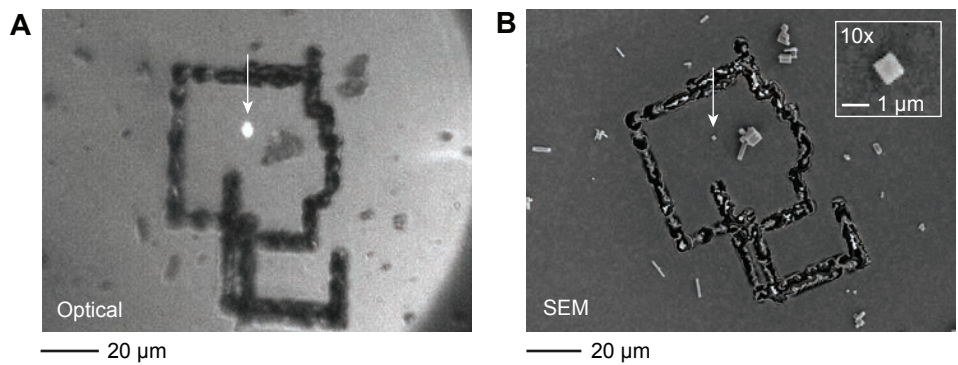

**Fig. S4. Marking samples for SEM. A**, Optical image of a laser device sample (arrow) after marking its position by deliberately creating damages on the substrate using high power laser with a specific pattern. The fluorescence from the crystal is superimposed, showing the crystal (arrow) as a bright spot. **B,** SEM image of the corresponding sample (arrow) identified by the particular laser damage pattern device. Inset, 10x magnified image of the CsPbBr$_3$ gain crystal. The crystal side length $L$ is measured to be 0.8.

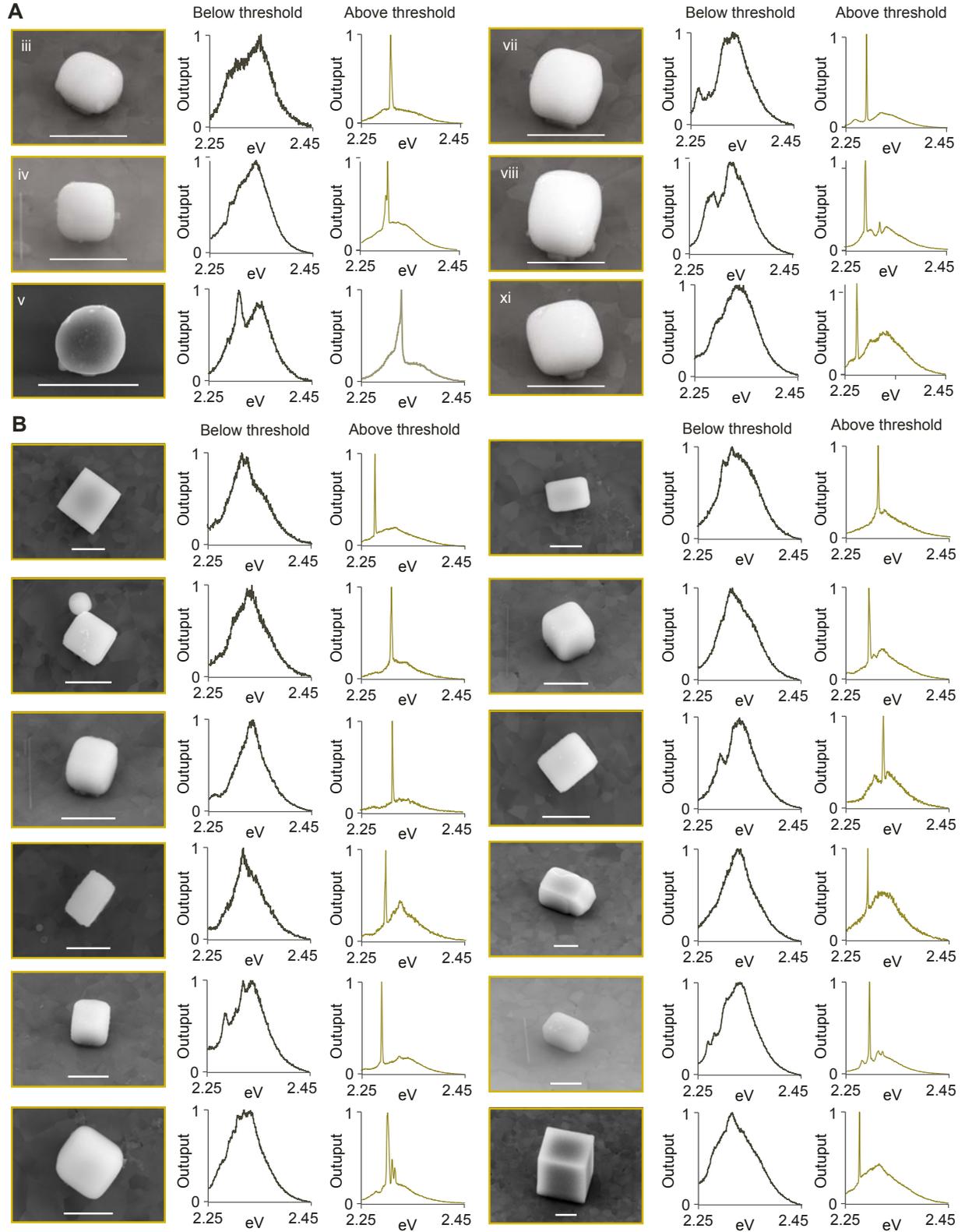

**Figure S5. Shape and spectra of plasmonic laser devices. A**, Submicron devices (*iii* to *xi* shown in Fig. 1B). **B**, microdevices. For each device, a representative SEM image (left), fluorescence emission (middle) below threshold (at $E = 0.2\, E_{th}$), and output spectrum (right) above lasing threshold (at $E = 1.2\, E_{th}$) are displayed. Scale bars, 1 μm.

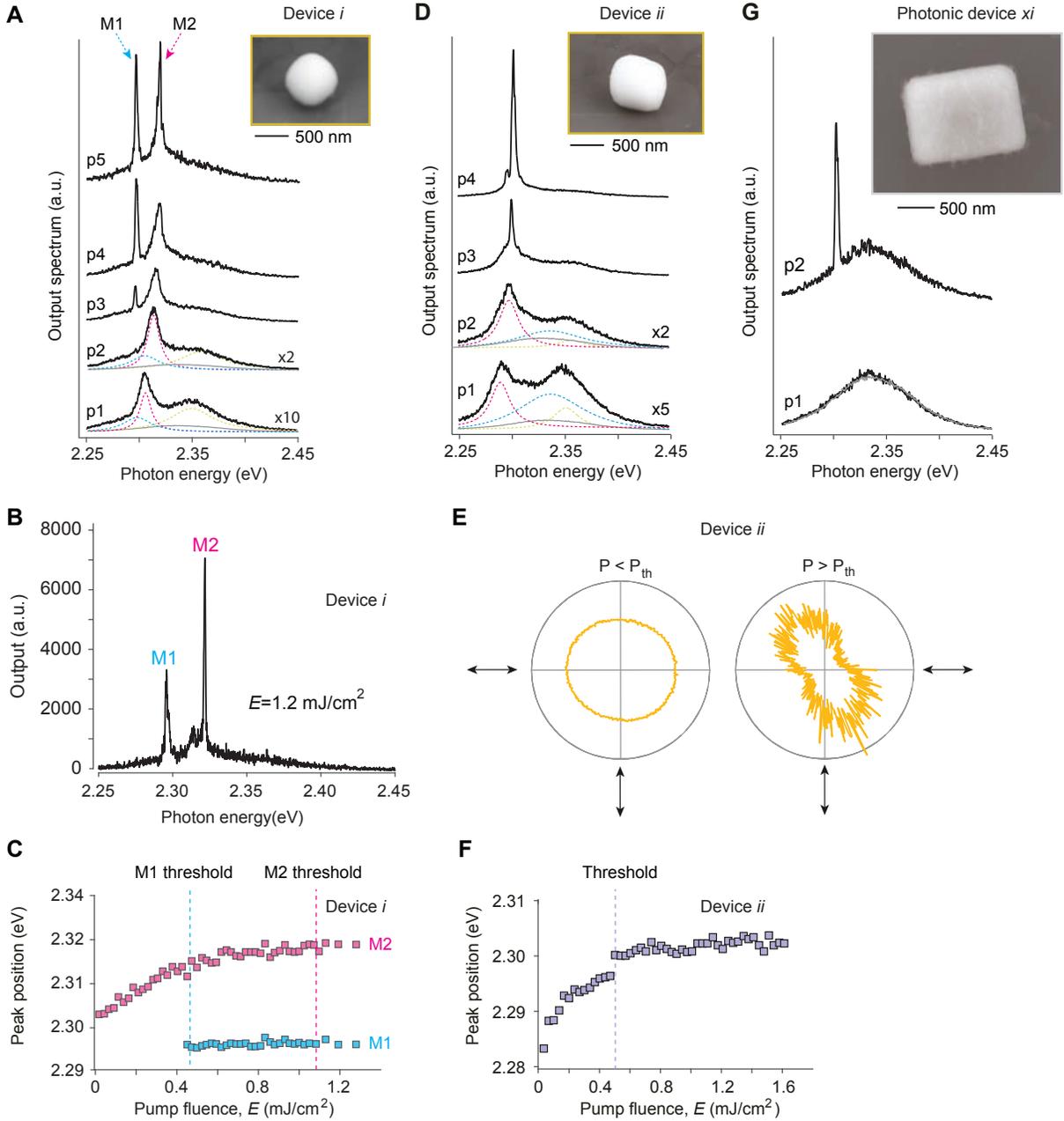

**Figure S6. Analysis of output spectra from submicron plasmonic devices.** **A-C**, Device *i*. **A**, PL spectra and SEM image (inset) of the smallest plasmonic laser *i* (*L*=0.57 μm) at various pump levels: p1 = 0.18, p2 = 0.33, p3 = 0.6, p4 = 0.88, and p5 = 1.1 mJ/cm$^2$. The spectra was fitted with multi-peak Lorentzian functions and typical fluorescence profile of CsPbBr$_3$. The number of Lorentzians (*N*) was decided using an analytical ray tracing model (See fig. S12). Two lasing modes are assigned as 'M1' (cyan curves) and 'M2' (magenta curves). Light green curve: non-lasing mode, Grey curve: typical fluorescence profile of CsPbBr$_3$. **B**, A representative output spectrum from a single pump pulse at *E*=1.2 mJ/cm$^2$ showing lasing of both modes. **C**, Spectral peak of the lasing modes (M1 and M2) as a function of pump fluence. **D-F**, Device *ii*. **D**, PL spectra from device *ii* (*L*= 0.61 μm) at various pump levels: p1 = 0.24, p2 = 0.57, p3 = 0.83, and p4 = 1.38 mJ/cm$^2$. **E**, Polar plots represent the emission intensity of the spontaneous emission below the threshold and the lasing mode above the threshold as a function of the polarization direction. **F**, Spectral peak of the mode. Dashed lines represent the threshold of the mode. **G**, PL spectra from the photonic device *xi* (*L*=1.3 μm).

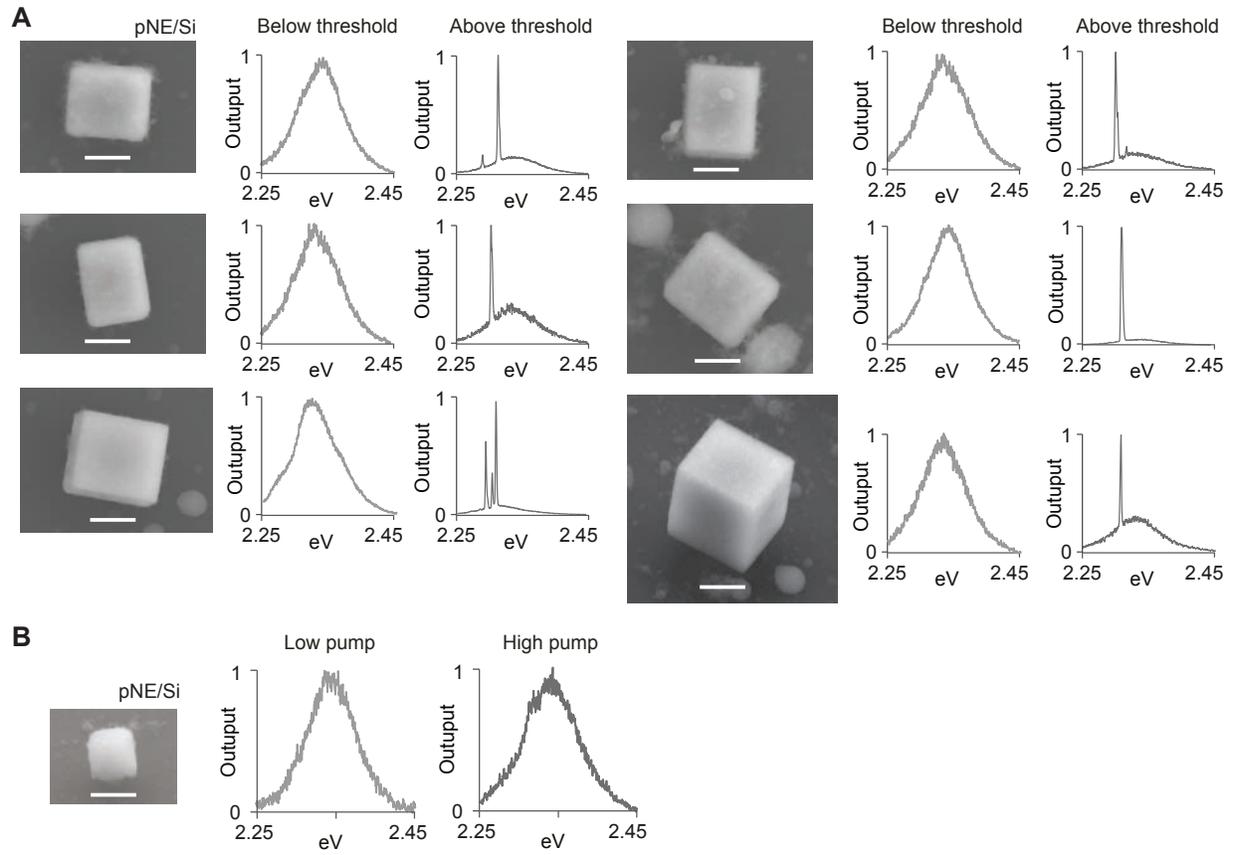

**Fig. S7. Shape and spectra of photonic devices. A**, Lasing photonic micro-lasers. For each device, a representative SEM image (left), fluorescence emission (middle) below threshold (at $E = 0.2\ E_{th}$), and output spectrum (right) above lasing threshold (at $E = 1.2\ E_{th}$) are displayed. **B**, A non-lasing submicron photonic device. An exemplary, non-lasing submicron photonic device. Mode structures are apparent in fluorescence emission spectra at both modest (< 0.5 mJ/cm²) and high (> 2 mJ/cm²; maximum possible without damaging the gain crystal) pump levels. Scale bar, 1 μm.

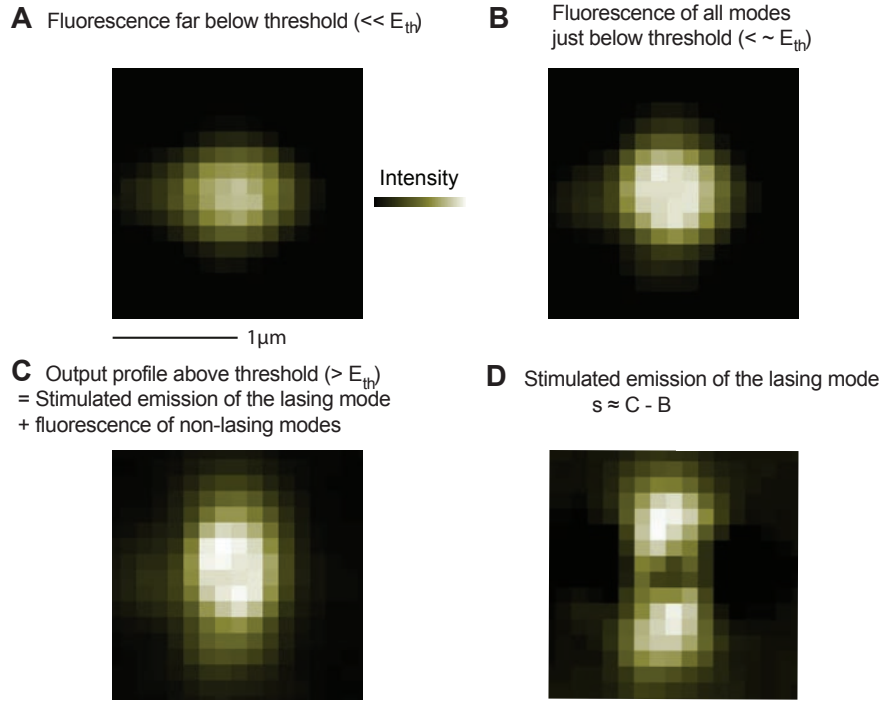

**Fig. S8. Reconstruction of the stimulated emission profile of the lasing mode.** **A**, Fluorescence image of the smallest plasmonic device *i* recorded in EMCCD (see the setup Fig. S3) at a very low pump level. **B**, Emission profile just below lasing threshold. **C**, The output profile of the device above threshold, which consists of the lasing mode and other non-lasing modes. **D**, Reconstructed emission profile of the lasing mode above threshold, obtained by subtracting **B** from **C**. This stimulated emission image is displayed in Fig. 2D.

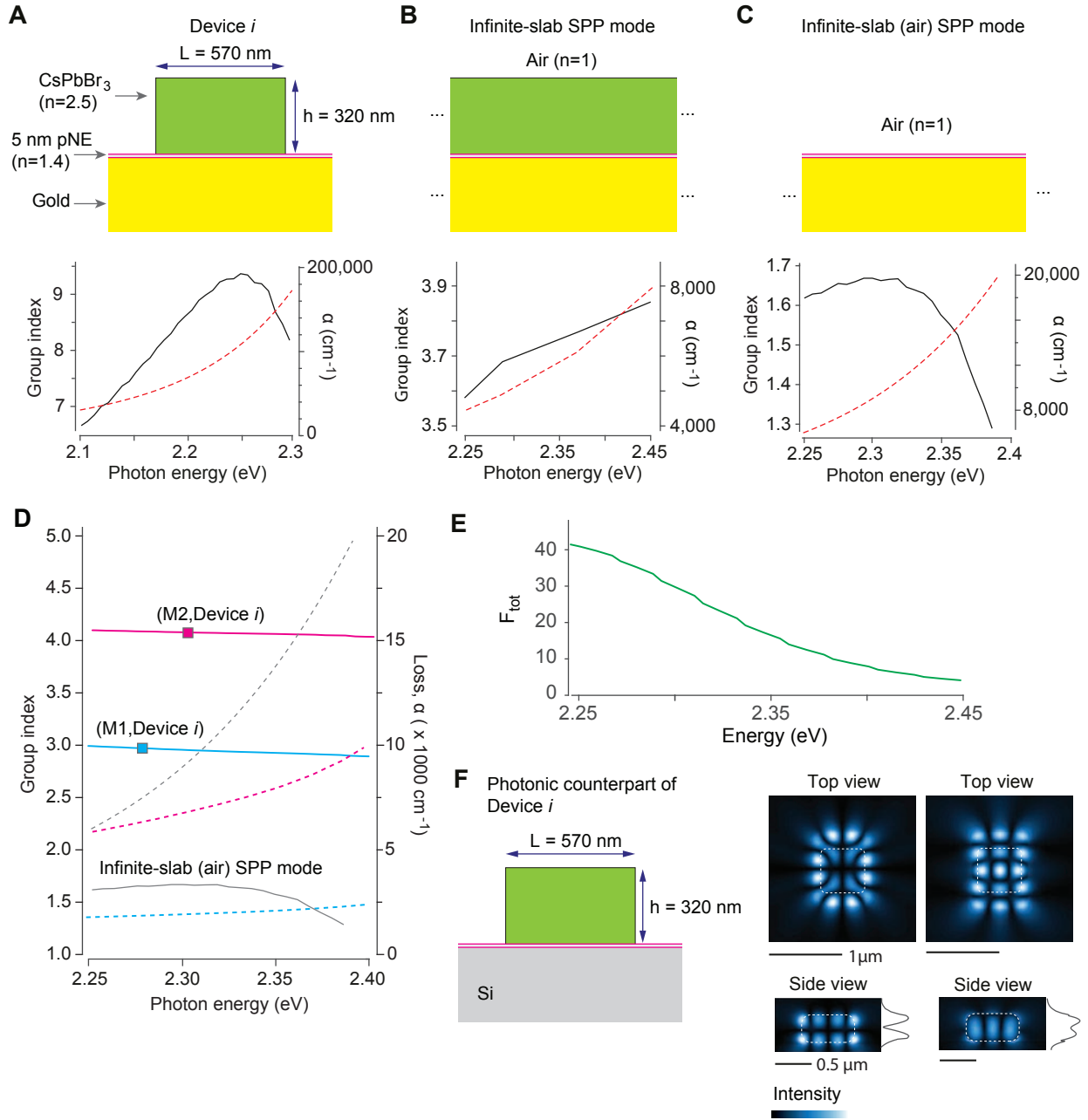

**Fig. S9. FDTD calculation results. A**, Dispersion of a pure-SPP mode supported in the plasmonic device *i* (Fig. 2A). **B**, Dispersion of a SPP wave guided in the perovskite-gold interface in a model of semi-infinite perovskite-pNE-gold slab. **C**, Dispersion of a SPP wave guided in the air-gold interface in a semi-infinite pNE-coated gold substrate. Solid lines (black): group index of the SPP waves, dashed line (red): propagation loss. **D**, Group refractive indices and absorption loss coefficients of the two lasing modes, 'M1' (cyan) and 'M2' (magenta), in the plasmonic device *i* and infinite-slab SPP mode, presented in **C** (gray). **E**, FDTD simulation of radiative Purcell factor $F_{tot}$ for a plasmonic device with $L = 0.8$ μm. **F**, FDTD result for a photonic laser model equivalent to the device *i* (i.e. exactly the same except that gold is replaced by Si). Two photonic modes are shown with energy at 2.34 eV and 2.28 eV, respectively. Despite the high real-part index of Si ($n = 4.1 + i\ 0.05$ at 2.3 eV), the gain medium supports WGMs propagating in the plane parallel to the perovskite-silicon interface. The Q-factors of these modes are similar to those in photonics devices on silica substrates ($n = 1.46$).

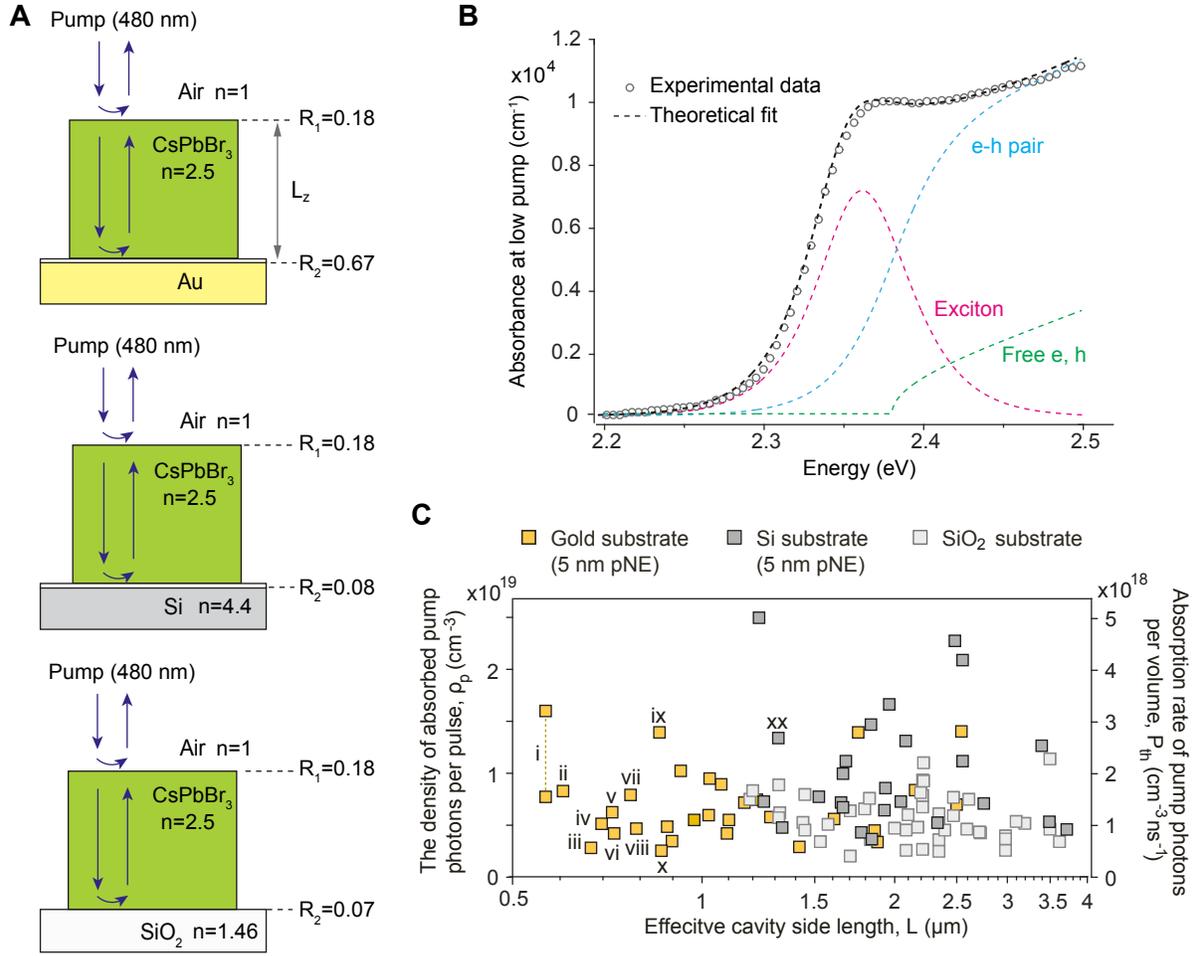

**Fig. S10. Pump absorption. A**, Ray-optic illustration of the pump light paths for the cases of gold, Si, and SiO₂ substrates, respectively. The ray-optic analysis is good approximation when the size of the gain medium is > 0.5 μm, in which case the gain medium can be considered a multimode waveguide for the pump light. **B**, Absorption spectra of CsPbBr₃ microcrystals. Circles, experimental data. Dashed curves: theoretical fits based on Elliot's theory (*1*) and Saha-Langmuir equation (*2*) for excitons with binding energy of 25 meV (magenta), electron-hole pairs (cyan), free electrons and holes (green), and a sum of these three components (Black) using Elliott's theory. The band gap energy is 2.38 eV. **C**, Threshold pump absorption for the three types of lasers, computed from the experimentally measured threshold pump fluence ($E_{th}$) (Fig. 1B) by taking into account pump reflection at the substrate.

<u>Methods</u>: The density of absorbed pump photons, $\rho_p$, at threshold can be expressed as: $\rho_p = \zeta E_{th}/(L_z \hbar\omega_p)$, where $L_z$ is the height of the gain medium, $\hbar\omega_p (= 4 \times 10^{-19}$ J) is the energy of single pump photon, and $\zeta$ is a conversion factor related to the reflection and absorption of the pump light. The conversion factor is shown to be (*3*): $\zeta = (1 - R_1) - [R_2(1 - R_1)e^{-kL_z} + (1 - R_2)](1 - R_1)e^{-kL_z}/(1 - R_1 R_2 e^{-2kL_z})$, where $R_1$ and $R_2$ are reflectivity, $k$ is the absorption coefficient of CsPbBr₃ for the pump wavelength ($10^4$ cm⁻¹ at 293 K) (*4*). For a CsPbBr₃ cube with a height of 1 μm, a total of 52% and 55% of the pump light is absorbed in plasmonic and photonic devices, respectively. The total absorbed pump photons can be converted to the rate of absorption of photon pumps per volume, $P_{th}$. In our experiments, the pump light is nanosecond pulses with a pulse duration of $\tau_p$. Then, the time-averaged absorption rate is given by: $P_{th} = \rho_p/\tau_p$. For example, for a 1 μm-high plasmonic laser device that reaches its lasing threshold at $E_{th} = 0.46$ mJ/cm², we get $\rho_p = 6 \times 10^{18}$ cm⁻³ and $P_{th} = 1.2 \times 10^{18}$ cm⁻³ ns⁻¹.

## Supplementary Note 1. Rate equations for a three-level laser system

### A. Laser rate equations

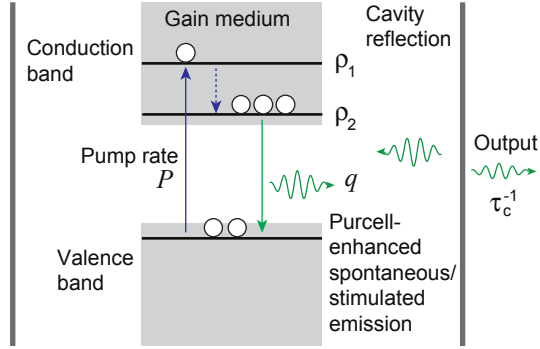

**Fig. S11. Schematic of the laser model**

The following rate equations can be written for a semiconductor laser (Fig. S11) (*5, 6*):

$$\frac{d}{dt}\rho_1(t) = P(t) - \frac{1}{\tau_{th}}\rho_1(t) \qquad (1)$$

$$\frac{d}{dt}\rho_2(t) = \frac{1}{\tau_{th}}\rho_1(t) - \frac{\beta_m V F_{tot}}{\tau_s}[\rho_2(t) - \rho_{tr}]q(t) - \frac{F_{tot}}{\tau_s}\rho_2(t) - \frac{1}{\tau_{nr}}\rho_2(t) \qquad (2)$$

$$\frac{d}{dt}q(t) = \frac{\beta_m V F_{tot}}{\tau_s}[\rho_2(t) - \rho_{tr}]q(t) + \frac{\beta F_{tot}}{\tau_s}\rho_2(t) - \frac{1}{\tau_c}q(t) \qquad (3)$$

where $\rho_1(t)$ and $\rho_2(t)$ denote the carrier density in the first and second excited states, respectively, $P(t)$ is the pump rate, and $\tau_{th}$ is the rate of non-radiative thermal decay from the first to the second excited levels, $\rho_{tr}$ is the number of carriers needed to reach transparency, $q(t)$ is the photon number density in the cavity, $F_{tot}$ is the total enhancement factor of emission contributed by all cavity modes including non-lasing modes as well as background modes uncoupled to a cavity: i.e. $F_{tot}=\sum F_i + F_u$, where $F_i$ is the enhancement factors of individual cavity modes and $F_u$ is the enhancement factor for modes uncoupled to the cavity ($F_u = 1$ when carriers are in free space, and $F_u > 1$ when the emitters are interacting with plasmonic waves that have high-density of states and are non-resonant in the cavity), $\beta_m$ is the spontaneous emission factor denoting the fraction of spontaneous emission into a cavity mode of interest and is related to the Purcell factor $F_m$ of the mode ($F_m = \beta_m F_{tot}$), $\tau_{th}$ is the thermalization lifetime, $\tau_s$ is the radiative lifetime, $\tau_{nr}$ is the non-radiative lifetime, $\tau_c$ is the photon lifetime inside the cavity.

Quantum yield η is defined as the fraction of radiative emission rate over a total decay rate:

$$\eta \equiv \frac{F_{tot}/\tau_s}{F_{tot}/\tau_s + 1/\tau_{nr}} = (1 + \frac{\tau_s}{F_{tot}\tau_{nr}})^{-1} \qquad (4)$$

### B. Lasing threshold

For pumping with a duration $\tau_p$ longer than $\tau_c$, $\tau_{th}$, and $\tau_s/F_{tot}$ (this condition is satisfied in our plasmonic lasers), we can set $\frac{d}{dt} = 0$ to obtain quasi-steady state solutions.:

$$P = \frac{\beta_m F_{tot} V}{\tau_s}(\rho_2 - \rho_{tr})q + \frac{F_{tot}}{\tau_s}\rho_2 + \frac{1}{\tau_{nr}}\rho_2 \qquad (5a)$$

$$0 = \frac{\beta_m F_{tot} V}{\tau_s}(\rho_2 - \rho_{tr})q + \frac{\beta F_{tot}}{\tau_s}\rho_2 - \frac{1}{\tau_c}q \qquad (5b)$$

The solution of $q$ can be expressed, using $p \equiv P/P_{th}$, as:

$$q(p) = \tau_c P_{th} \frac{(p-1) + \sqrt{(p-1)^2 + 4\beta_m \eta p}}{2} \quad (6)$$

$$P_{th} V = \frac{1}{\beta_m F_{tot} \tau_c} \left( F_{tot} + \frac{\tau_s}{\tau_{nr}} \right) + \frac{\rho_{tr} V}{\tau_s} \left( F_{tot} - \beta_m F_{tot} + \frac{\tau_s}{\tau_{nr}} \right) \quad (7)$$

$q$ as a function of $P$ has a nonlinear kink at $p = 1$, so this point defines a lasing threshold, and $P_{th}$ is the pump rate at threshold.

The second term in (7) is smaller than the first term in plasmonic lasers with $V < 1$ μm³ (or $L < \sim 1$ μm). This can be shown using $\rho_{tr} = \sim 10^{17}$ cm⁻³ (7) and our experimental values: $F_{tot} = 60$, $\tau_s = 300$ ns, $\tau_c = 30$ fs, and $\beta_m = 0.1$.

Neglecting the second term and using the quality factor of the cavity mode, $Q_m = \omega_0 \tau_c$, we derive:

$$P_{th} V \approx \frac{\omega_0}{Q_m \cdot \beta_m \cdot \eta} \quad (8)$$

Plotting (6) in the log-log scale, a nonlinearity parameter $x$ can be defined as

$$x \equiv \beta_m \cdot \eta \quad (9)$$

### C. Threshold carrier density

One of the reasonable definitions of lasing threshold is the point when a stimulated emission rate ($\frac{\beta_m F_{tot} V}{\tau_s} \rho_2 q$) is equal to the spontaneous emission rate ($\frac{\beta_m F_{tot}}{\tau_s} \rho_2$) into the lasing mode. This condition is equivalent to one photon inside the cavity, i.e. $qV = 1$. In this condition, from Eq. (5a) we find:

$$P_{th} = \frac{\beta_m F_{tot}}{\tau_s}(\rho_{th} - \rho_{tr}) + \frac{1}{\eta} \frac{F_{tot}}{\tau_s} \rho_{th} \approx \frac{1}{\eta} \frac{F_{tot}}{\tau_s} \rho_{th} \quad (10)$$

where $\rho_{th}$ is the carrier density at threshold, and the approximation is valid because $\rho_{th} - \rho_{tr} < \rho_{th}/\beta_m \eta$.

From (7) and (10), we find:

$$\rho_{th} V \approx \frac{\omega_0 \tau_s}{Q_m \cdot \beta_m \cdot F_{tot}} \quad (11)$$

As another definition of lasing threshold, it is the point when net stimulated emission rate, $(\beta_m F_{avg} V(\rho_2 - \rho_{tr})/\tau_s)$, is equal to the cavity loss rate ($1/\tau_c$). At this condition, we get:

$$\rho_2 - \rho_{tr} = \frac{\omega_0 \tau_s}{Q_m \cdot \beta_m \cdot V \cdot F_{tot}} \quad (12)$$

Since $\rho_2 - \rho_{tr} \approx \rho_2$ in most cases, we find $\rho_{th} V \approx \frac{\omega_0 \tau_s}{Q_m \cdot \beta_m \cdot F_{tot}}$, which is consistent with (11). The net gain in the laser is approximately proportional to $\rho_{th} V$.

### D. Transient fluorescence decay profiles

Consider fluorescence measurement using ultrashort pump pulses with a duration of $\tau_p$. At low pump fluence below lasing threshold, the stimulated emission term in (2) is neglible. Following the absorption of a single pump pulse at $t=0$, the carrier density and magnitude of fluorescence decay over time:

$$\rho_2(t) = P\tau_p e^{-\left(\frac{F_{tot}}{\tau_s} + \frac{1}{\tau_{nr}}\right)t} \quad (13)$$

$$\frac{d}{dt}q(t) = \frac{F_{tot}}{\tau_s} \rho_2 = P\tau_p \frac{F_{tot}}{\tau_s} e^{-\frac{t}{\tau_{tot}}} \quad (14)$$

where $\tau_{tot} = (\frac{F_{tot}}{\tau_s} + \frac{1}{\tau_{nr}})^{-1}$ is total decay lifetime. Integrating (14) over time yields.

$$q(t) = P\tau_p \tau_{tot} \frac{F_{tot}}{\tau_s}\left(1 - e^{-\frac{t}{\tau_{tot}}}\right) \quad (15)$$

The total number of fluorescence photons collected is given by

$$q(\infty) = \eta \cdot P\tau_p \quad (16)$$

where $\eta = (1 + \frac{\tau_s}{F_{tot}\tau_{nr}})^{-1}$ was used. For given pump pulse parameters, the total number of fluorescence photons is proportional to $\eta$, as expected from the definition of $\eta$.

From (15), we find

$$\frac{d}{dt}q(t)|_{t=0}| = F_{tot}P\tau_p/\tau_s \quad (17)$$

For given pump pulses and intrinsic radiative time constant, the initial fluorescence peak is proportional to $F_{tot}$.

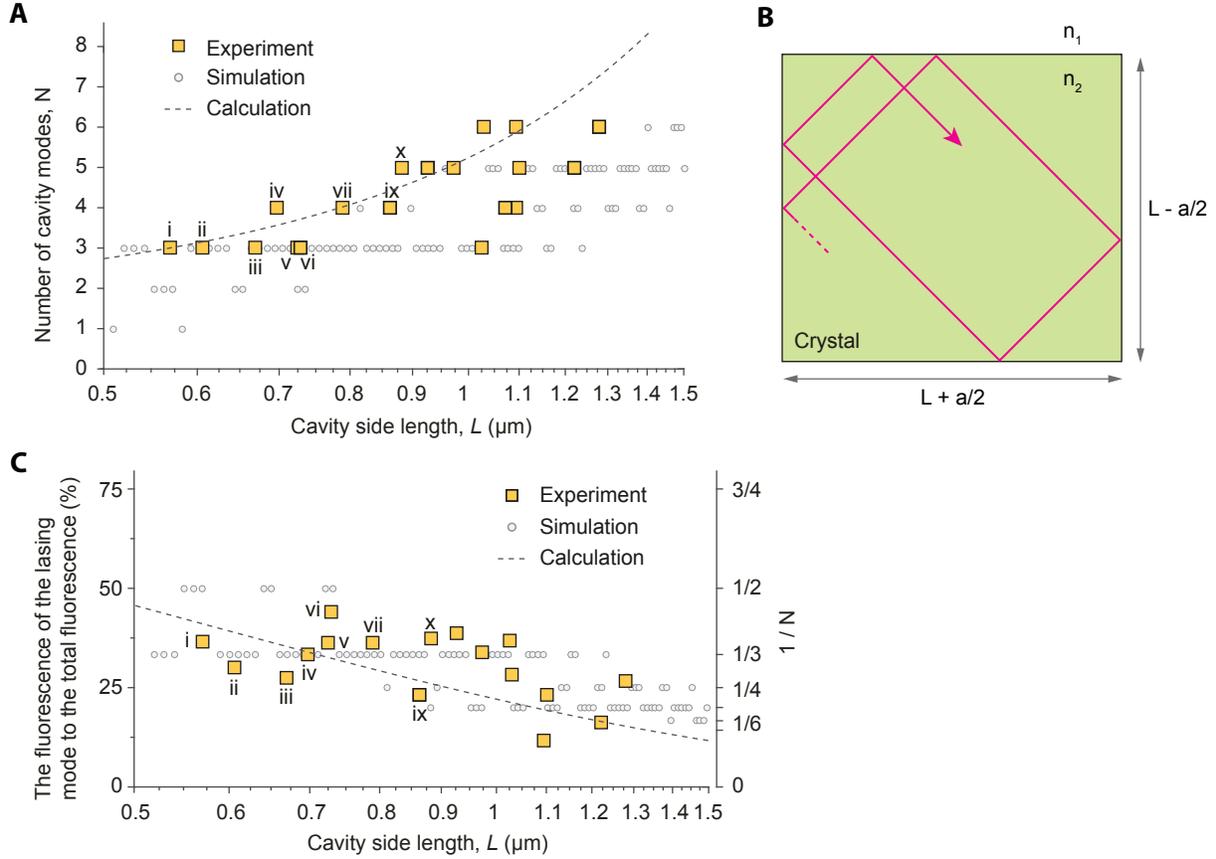

**Fig. S12. Number ($N$) of cavity modes. A**, The number of cavity modes $N$ in the plasmonic devices. Square: experimental data, circles: simulation result, and dashed line: analytic approximation. **B**, Schematic of the 2D rectangular cavity model. **C**, A plot of $1/N$. Square: the magnitude ratio of the fluorescence emission from the lasing mode (below threshold) over the total fluorescence, circles: simulation result for $1/N$, and dashed line: analytic approximation.

<u>Methods</u>: The resonance condition for cavity modes is given by $(m_x + m_y)\lambda = 2n[(L + \frac{a}{2})\cos\theta + [(L - \frac{a}{2})\sin\theta]$, where $m_x$ and $m_y$ are integers, $N = m_x + m_y$, $n$ is the effective refractive index of the modes in the crystal, and $\theta$ is a beam angle to the surface normal within a range between $\theta_c$ and $\pi/2 - \theta_c$, where $\theta_c = \sin^{-1}(1/n)$ is critical angle for total internal reflection (8). The circles in **A** and **C** were generated for several different aspect ratios ($a/L = 1, 1.05, 1.1, 1.15,$ and $1.2$). From the spread of these simulation data, the variability observed in the experimental data can be attributed to small differences in the aspect ratio and shape of the gain crystals in the devices. An approximation solution of the cavity resonance condition has been derived (9): $N \approx Round\ [32 n_g n \frac{\delta v}{v} (\frac{L}{\lambda})^2 (1 - \sin(\pi/4 + \theta_c))]$, where $n_g$ is group refractive index, and $v$ and $\delta v$ are the center frequency and full-width-half-maximum (FWHM) of the optical gain spectrum, respectively. This formula was used to plot the dashed curve in **A** and **C**, with reasonably good correspondence to the simulation and experimental results.

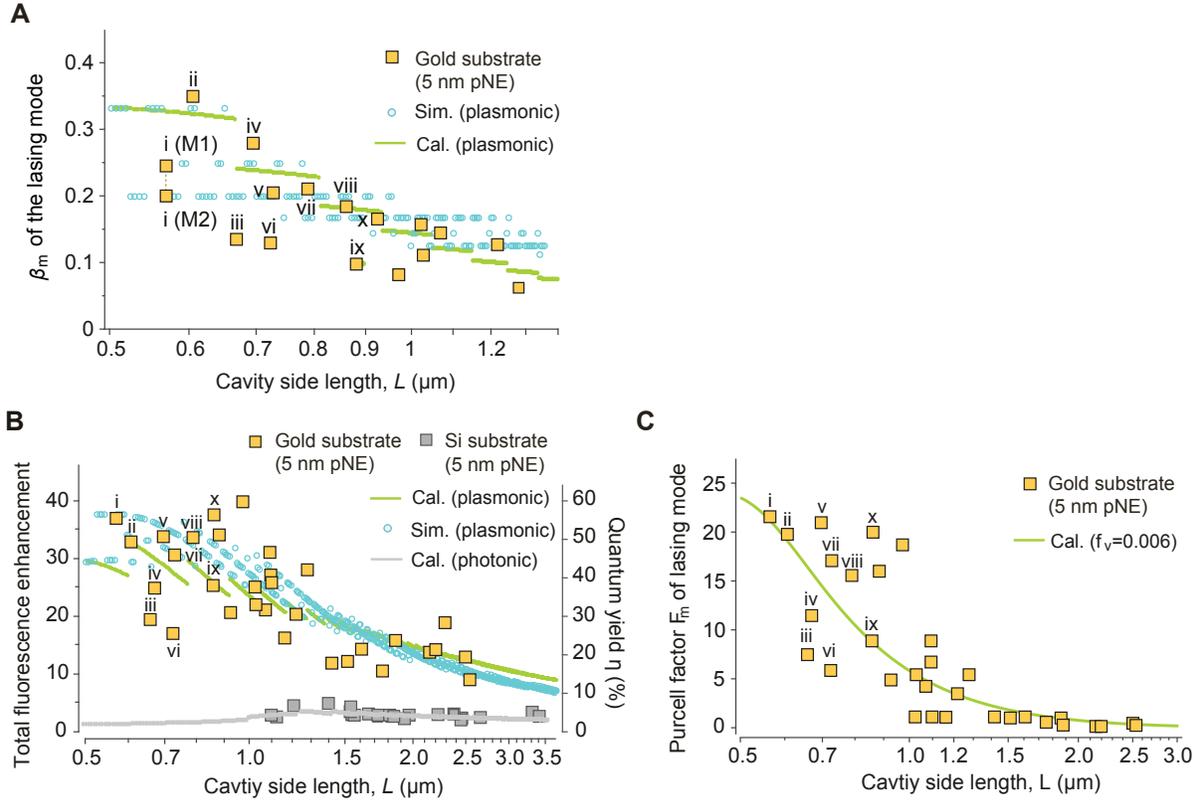

**Fig. S13. Analytic model tracking the general trends of $\beta_m$, $\eta$ and F$_m$. A**, The spontaneous emission factor of the lasing mode $\beta_m$ for the plasmonic devices (yellow squares) and a analytical model (green curves) and a simulation result (cyan circles). **B**, Fluorescence intensity measured from plasmonic (yellow) and photonic (grey) devices, and their quantum yields $\eta$ estimated rom the data relative to reference, larger (> 5 μm, $\eta \approx 1.5\%$) CsPbBr$_3$ crystals on Si (data not shown). Curves: analytic calculations. Circles: simulation result. **C**, Purcell factor of the lasing mode, $F_m$, of the plasmonic lasers (Squares: experimental data; curve, analytic fit).

<u>Methods:</u> We used theoretical models to calculate $\beta_m = F_m/F_{tot}$, $\eta = (1 + \frac{\tau_s}{F_{tot}\tau_{nr}})^{-1}$, and $F_m = \frac{3\lambda^3}{4\pi^2}\frac{Q_m}{V_m}$ where $Q_m$ was calculated for WGM in spherical cavities with a diameter $L$, and $V_m = f_V V$ is mode volume. Volumetric factor $f_V$ was the fitting parameter. For plasmonic devices (assuming $F_u = 5$), we found $f_V \approx 0.006$, corresponds to a modal height of 3-10 nm. For photonic devices ($F_u = 1$), $f_V = 0.9$ produced the best fit. The overall quality of fitting for $\beta_m$, $\eta$, and $F_m$ achieves descent agreement with the experimental data considering the simple assumptions (e.g. same $V_m$ and $Q_m$ for all modes.)

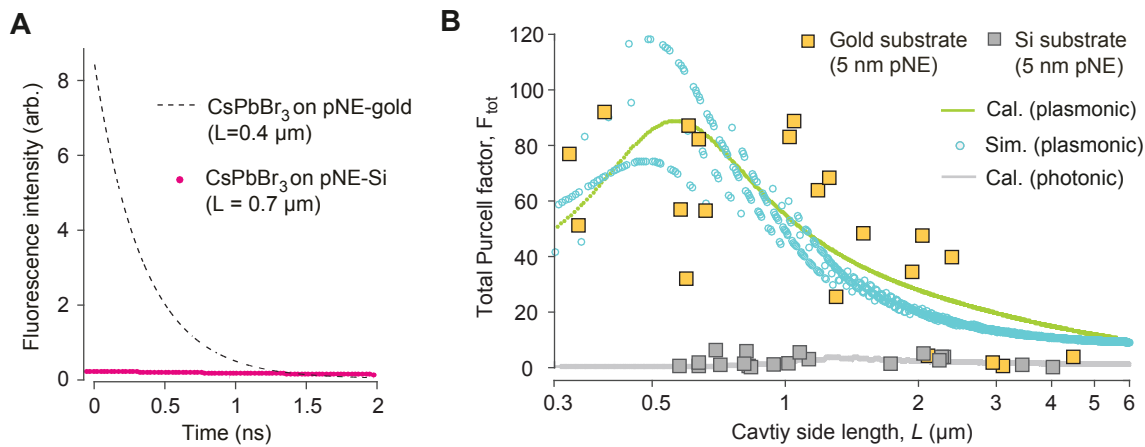

**Fig. S14. Purcell factors measured by lifetime measurement. A**, Transient PL curves of a photonic CsPbBr₃ submicron crystals on pNE-Si substrates (magenta circles) and a plasmonic crystal on pNE-gold (dashed line, Fig. 3D). **B**, Total Purcell factor, $F_{tot}$, from 23 CsPbBr₃ samples on gold and 20 samples on Si, in reference to the mean value of large CsPbBr₃ crystals (dominantly $L > 5$ μm) on Si substrates. Note that lasing experiments were not performed for these samples, so some of these samples may be non-lasing even at high pump levels (> 2 mJ/cm²). Squares: experimental data; Solid curves: analytic calculation, Circles: simulation result. Both analytic calculation and simulations were computed using $f_v$ = 0.006 for plasmonic devices (green) and $f_v$ = 0.9 for photonic devices (grey).

**Table S1. Time constants measured by transient fluorescence decay analysis**

|  | Plasmonic CsPbBr$_3$ on pNE-gold | | | Photonic CsPbBr$_3$ on pNE-Si | | |
|---|---|---|---|---|---|---|
|  | $L$ = 0.4 μm | $L$ = 0.9 μm | $L$ = 3.8 μm | $L$ = 0.7 μm | $L$ = 2.4 μm | $L$ = 4.3 μm |
| $\tau_1$ from TCSPC* | 0.7 ns | 1.2 ns | 2.6 ns | 4.1 ns | 2.5 ns | 3.2 ns |
| $\tau_2$ from TCSPC | - | - | 18 ns | - | 5.8 ns | 5.6 ns |
| $A_1$ from TCSPC | 1 | 1 | 0.76 | 1 | 0.62 | 0.68 |
| $A_2$ from TCSPC | - | - | 0.23 | - | 0.38 | 0.32 |
| $\tau_{tot}$ measured | 0.7 ns | 1.2 ns | 3.3 ns | 4.1 ns | 3.8 ns | 3.7 ns |
| $F_{tot}$ measured | 50 | 20 | 1 | 0.5 | 5 | 1 |
| $\eta$ measured** | 0.4 | 0.1 | 0.015 | 0.007 | 0.075 | 0.015 |
| $\tau_s$ computed | 88 ns | 240 ns | 220 ns | 270 ns | 250 ns | 250 ns |
| $\tau_{nr}$ computed | 0.71 ns | 1.2 ns | 3.3 ns | 4.1 ns | 3.8 ns | 3.7 ns |

* Time-correlation single photon counting (TCSPC). Fluorescence lifetime was measured from exponential fit to the fluorescence decay curves using $\tau_{tot} = A_1\tau_1 + A_2\tau_2$.

**To determine baseline $\eta$ of CsPbBr$_3$, we used as-made CsPbBr$_3$ microcrystals (dominantly with $L > 5$ μm) in DMF solution. The samples were excited by using a picosecond frequency doubled laser ($\lambda$ = 382 nm) at low pumping ($\rho_p$ = ~10$^{16}$ cm$^{-3}$). The total magnitude of output fluorescence was measured. This apparatus was calibrated with respect to 10 μM of fluorescein dye in 0.1 N of NaOH aqueous solution (*10*), which has a known $\eta$ of 0.92. The measured intrinsic $\eta$ of the CsPbBr$_3$ samples was 0.015.

## Supplementary Note 2. Calculation used to plot the theoretical curves in Fig. 4

The total Q-factor, $Q_{tot}$, of a cavity consists of radiative Q-factor, $Q_{rad}$, and absorptive Q-factor $Q_{abs}$: $1/Q_{tot} = 1/Q_{rad} + 1/Q_{abs}$. The radiative Q-factor of WGM in a spherical cavity using the analytic equation was calculated using an approximate formula (*11*).

The large group index and stronger surface reflection of plasmonic modes in the device boundary makes $Q_{rad}$ for plasmonic lasers much higher than $Q_{rad}$ in photonic devices. On the contrary, high metallic absorption (with a coefficient of $\alpha$) in plasmonic devices makes $Q_{abs}$ in photonic devices is much lower than $Q_{abs}$ in photonic devices. In submicron devices, these effects approximately cancel each other: $Q_{tot}$ in plasmonic and photonic devices are about the same, according to experiments and FDTD calculations, within a factor of ~2.

To calculate $Q_{rad}$, we used the well-known WGM mode theory using a core group index of 3.8, a clad index of 1.65, and absorption coefficient $\alpha$ = 5,800 cm$^{-1}$ for plasmonic lasers, and a core index of 2.0, a clad index of 1, and an absorption coefficient $\alpha$ = 600 cm$^{-1}$ for photonic lasers. The absorption coefficient of photonic lasers was adapted from the measured absorption spectra of CsPbBr$_3$ microcrystals presented in Fig. S10B.

## Supplementary Note 3. Critical carrier density for the Mott transition to electron hole plasma (EHP)

At a very low carrier density, electrons and holes in CsPbBr$_3$ are strongly correlated via Coulomb interaction (*2*). As the carrier density increases, the mean distance between them decreases, and when it is less than a critical distance, electromagnetic screening become so significant that the Coulomb interaction becomes negligible. In this electron-hole plasma (EHP) state, the active carriers behave like free charges without Coulomb interaction. The phase transition to EHP is called the Mott transition. Let the critical transition density $\rho_{Mott}$. A model developed by Haug and Schmitt-Rink(*12*) estimates $\rho_{Mott} \approx 0.028(k_B T_e/(E_X a_B{}^3))$. Here, $k_B$ is Boltzmann constant, $T_e$ (= 709 K; see Fig. S15) is electronic temperature, $a_B$ is Bohr radius — $a_B = a_H \varepsilon_r^X m_e/\mu \approx 3.5$ nm for CsPbBr$_3$, where $a_H$ = 0.053 nm, $\mu$ (= 0.125) is the reduced mass of electron-hole pair (*13*), $m_e$ is electron mass, and $\varepsilon_r^X = \sqrt{(13.6\ eV/E_X)\mu/m_e}$ is the dielectric function of Coulomb-correlated carriers — and $E_X \approx 25$ meV is exciton binding energy for CsPbBr$_3$. We find $\rho_{Mott} \approx 9\times10^{17}$ cm$^{-3}$ for CsPbBr$_3$. In experiments, the carrier density at lasing threshold exceeds $10^{18}$ cm$^{-3}$ at which the EHP state is established.

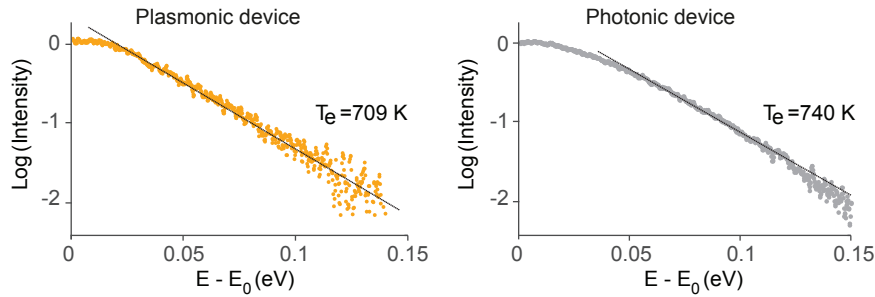

**Fig. S15. Measurement of $T_e$ for the calculation of $\rho_{Mott}$.** The blue-wing tail of PL spectra at $P = 0.9\ P_{th}$ is shown for a plasmonic (left) and a photonic (right) device. The fitting curve was the classical Maxwell-Boltzmann distribution function, which describes the EHP state(*14*): $N(E) \propto \exp(-(E - E_0)/k_B T_e)$, where $N(E)$ ($\propto I$) is the number density of carriers with energy $E$ and $E_0$ is the electronic state energy at the peak PL, and $T_e$ is electronic temperature. We found $T_e$ to be 709 and 740 K, respectively.

**Table S2. A list of representative visible plasmonic nanolasers demonstrated to date adopted from recent review article (*15*).**

| Type | Material (λ) | Gain-medium size (**Largest dimension**) | $\lambda_p$ | Year (Ref) |
|---|---|---|---|---|
| Cubes | CsPbBr$_3$/Au (540 nm) | Length: **600 nm**, Height: 400 nm | 5 ns | This work |
| Nanowires | ZnO/Al (380 nm) | Length: **1-4 μm**, Diameter: 70 nm | 0.5 ns | 2016 (*16*) |
| Nanowires | GaN/Al (375 nm) | Length: **15 μm**, Diameter: 100 nm | 10 ns | 2014 (*17*) |
| Plates | CdS/Ag (500 nm) | Length: **1 μm**, Diameter: 45 nm | 100 fs | 2011 (*18*) |